\documentclass[12pt]{article}
\pdfoutput=1
\usepackage{hyperref}
\usepackage{graphicx}
\usepackage{graphics}
\usepackage{dsfont}
\usepackage{epsfig}
\usepackage{amsmath,amssymb,amsthm,amscd}

\newcommand{\Tr}{\textrm{Tr}}

\def\ket{\rangle}
\def\bra{\langle}

\newcommand{\be}{\begin{equation}}
\newcommand{\ee}{\end{equation}}
\newcommand{\ba}{\begin{aligned}}
\newcommand{\ea}{\end{aligned}}

\numberwithin{equation}{section}

\begin{document}
\begin{titlepage}

\rightline{USTC-ICTS/PCFT-21-18}

\vskip 3 cm

\centerline{\Large 
\bf  
Non-Negativity of BMN Two-Point Functions } 
\vskip 0.2 cm
\centerline{\Large 
\bf 
 With Three String Modes  }

\vskip 0.5 cm

\renewcommand{\thefootnote}{\fnsymbol{footnote}}
\vskip 30pt \centerline{ {\large \rm 
Bao-ning Du\footnote{baoningd@mail.ustc.edu.cn}, 
Min-xin Huang\footnote{minxin@ustc.edu.cn}  
} } \vskip .5cm  \vskip 20pt 

\begin{center}
{Interdisciplinary Center for Theoretical Study,  \\ \vskip 0.1cm  University of Science and Technology of China,  Hefei, Anhui 230026, China} 
 \\ \vskip 0.3 cm
{Peng Huanwu Center for Fundamental Theory,  \\ \vskip 0.1cm  Hefei, Anhui 230026, China} 
\end{center}

\setcounter{footnote}{0}
\renewcommand{\thefootnote}{\arabic{footnote}}
\vskip 40pt
\begin{abstract}

Recently, we proposed a novel entry of the pp-wave holographic dictionary, which equated the Berenstein-Maldacena-Nastase  (BMN)  two-point functions in free $\mathcal{N}=4$ super-Yang-Mills theory with the norm squares of the quantum unitary transition amplitudes between the corresponding tensionless strings in the infinite curvature limit, for the cases with no more than three string modes in different transverse directions. A seemingly highly non-trivial conjectural consequence, particularly in the case of three string modes, is the non-negativity of the BMN two-point functions at any higher genus for any mode numbers. In this paper, we further perform the detailed calculations of the BMN two-point functions with three string modes at genus two, and explicitly verify that they are always non-negative through mostly extensive numerical tests.

\end{abstract}

\end{titlepage}
\vfill \eject


\newpage

\baselineskip=16pt

\tableofcontents

\section{Introduction}

We continue the studies of free BMN correlators in our recent paper \cite{Du:2021dml}. The general motivations and the physical settings are explained in the previous paper. We shall provide a recapitulation with some new comments more relevant to the current context.    

The AdS/CFT correspondence \cite{Maldacena:1997re, Gubser:1998bc, Witten:1998qj} is a conceptual breakthrough in our understandings of quantum gravity, in particular provides a non-perturbative definition of string theory in AdS background in terms of $\mathcal{N}=4$ $SU(N)$ super-Yang-Mills theory. We consider the Penrose limit \cite{Penrose1976}, which gives rise to another maximally supersymmetric background \cite{Blau:2001ne}, known as the pp-wave or plane wave geometry 
\begin{eqnarray} \label{ppwave}
ds^2 = -4 dx^{+} dx^{-} -\mu^2 (\vec{r}^{~2} + \vec{y}^{~2} ) (dx^{+})^2 + d \vec{r}^{~2} + d  \vec{y}^{~2}, 
\end{eqnarray}
where $x^{+}, x^{-} $ are light cone coordinates, $\vec{r}, \vec{y}$ are 4-vectors, and the parameter $\mu$ measures the spacetime curvature as well as the Ramond-Ramond flux $F_{+1234}=F_{+5678}\sim \mu$. This appears to be a promising ground for quantitative explorations of the holographic duality in stringy regimes, as the dual theories on both sides can be either free or weakly coupled. 

In the groundbreaking paper \cite{Berenstein:2002jq}, Berenstein, Maldacena and Nastase  (BMN) proposed a type of near-BPS operators, which correspond to the type IIB closed strings on the pp-wave background. The free string spectrum is correctly reproduced by gauge interactions as the planar conformal dimensions of BMN operators. The BMN scaling limit with large R-charge $J\sim \sqrt{N}\sim \infty$ appears to be the right Goldilocks limit in this situation, since a smaller R-charge would not provide finite string interactions in the strict $N\sim \infty$ limit, while  a larger R-charge may blow up the strings into D-branes, known as giant gravitons, studied in early papers e.g. \cite{McGreevy:2000cw, Hashimoto:2000zp, Balasubramanian:2001nh, Corley:2001zk, Balasubramanian:2002sa, Balasubramanian:2004nb}. Some recent studies relating to the large R-charge limit or Penrose limit, as well as applications in more general theories can be found in e.g. \cite{Hellerman:2018xpi, Kemp:2019log, Gaume:2020bmp, deMelloKoch:2020jmf, Berenstein:2020jen, Dodelson:2020lal, Yang:2021kot}. 

As in our previous papers \cite{Huang:2002wf, Huang:2002yt, Huang:2010ne, Huang:2019lso, Huang:2019uue, Du:2021dml}, we focus on free gauge theory and study the BMN correlation functions. This corresponds to the pp-wave background with infinite curvature and infinite Ramond-Ramond flux as $\mu\sim \infty$ in the geometry (\ref{ppwave}), where the strings are tensionless with completely degenerate spectrum. There are still interesting string interactions as we identify the finite genus-counting parameter $g:=\frac{J^2}{N}$ as the effective string coupling constant in this case. Some non-planar BMN correlators are first computed in \cite{Kristjansen:2002bb, Constable:2002hw}. 

The celebrated standard AdS holographic dictionary \cite{Witten:1998qj} seems not directly applicable for string interactions in the pp-wave background, as the geometries are quite different. So in some cases, certain guessworks may be required to identify the correct entries of the ``pp-wave holographic dictionary".  In this paper we focus on a probability interpretation of BMN two-point functions \cite{Huang:2019lso, Du:2021dml}. There are other interesting entries of the pp-wave holographic dictionary, namely the comparisons of free planar BMN three-point functions with Green-Schwarz light-cone string field  cubic vertices \cite{Green:1982tc, Green:1984fu, Spradlin:2002ar, Huang:2002wf, Pankiewicz:2002tg}, the factorization formulas \cite{Huang:2002yt, Huang:2010ne}, which are most recently explored in the recent paper \cite{Du:2021dml} in the context of many string modes. Higher point correlators, including the planar three-point functions, actually always vanish in the strict BMN limit, and are now perceived by us as a kind of virtual processes.  

Let us introduce some notations. The BMN vacuum operator is simply proportional $\Tr(Z^J)$ where $Z$ is a complex scalar in the $\mathcal{N}=4$ $SU(N)$ super-Yang-Mills theory. One can insert the four remaining real scalars into the trace with phases, corresponding to string modes in four of the eight transverse dimensions. The BMN operators are then denoted as $O^J_{(m_1, m_2, \cdots, m_k)}$, where the positive and negative integer modes represent the left and right moving stringy excited modes, while the zero modes are supergravity modes representing discretized momenta in the corresponding traverse direction. We will consider the case of string modes in different transverse directions, otherwise the BMN operators are no longer near-BPS and there may be some potential issues as discussed in \cite{Du:2021dml}. Due to the closed string level matching condition $\sum_i m_i =0$, the excited stringy states have at least two string modes. The BMN operators are properly normalized to be orthonormal at planar level, and the genus $h$ two-point functions are proportional to $g^{2h}$ as
\be \ba
&\bra \bar{O}^J_{(m_1, m_2, \cdots, m_k)} O^J_{(n_1, n_2, \cdots, n_k) }\ket_0 = \delta_{m_1, n_1} \cdots  \delta_{m_k, n_k} , \\ 
& \bra \bar{O}^J_{(m_1, m_2, \cdots, m_k)} O^J_{(n_1, n_2, \cdots, n_k) }\ket_h\sim g^{2h} . 
\ea \ee

As discussed in \cite{Du:2021dml}, the BMN two-point functions are real and symmetric, and there is a nice normalization relation summing over one set of mode numbers 
\be \label{normalization}
\sum_{\sum_{i=1}^k n_k=0 }\bra \bar{O}^J_{(m_1,m_2,\cdots ,m_k)} O^J_{(n_1,n_2,\cdots ,n_k)} \ket_{h} = 
\frac{g^{2h} }{2^{2h} (2h+1)!}  . 
\ee
We may define a matrix element, summing up all genus contributions with a proper normalization by the all-genera formula of vacuum correlator
\be \label{matrixp}
p_{(m_1,m_2,\cdots ,m_k), (n_1,n_2,\cdots ,n_k)} = \frac{g}{2\sinh(\frac{g}{2})} \sum_{h=0}^{\infty} \bra \bar{O}^J_{(m_1,m_2,\cdots ,m_k)} O^J_{(n_1,n_2,\cdots ,n_k)} \ket_{h} , 
\ee
so that it looks like a probability distribution 
\be \label{normalization1}
\sum_{\sum_{i=1}^k n_k=0 } p_{(m_1,m_2,\cdots ,m_k), (n_1,n_2,\cdots ,n_k)}  =  1.  
\ee
To interpret the matrix elements as a  probability distribution, they need to be non-negative. For the case of two string modes, the non-negativity at any genus can be easily proven since the two string modes are opposite numbers \cite{Huang:2019lso}, while for the case of four string modes, it turns out that the genus one two-point functions can be negative \cite{Du:2021dml}. There seems to be a rule forbidding the ``crowdedness" of string modes, that we can not holographically use up all four remaining scalars to fully occupy the four transverse dimensions with $SO(4)$ rotational symmetry unbroken by the Ramond-Ramond flux in the pp-wave background. In this paper, we focus on the case of three string modes. Unlike the case of two string modes, we are not aware of a simple analytic proof of the non-negativity. Instead, we perform the detailed calculations and explicitly verify the non-negativity up to genus two. 

The normalization relations and non-negativity with two and three string modes suggest a novel entry of the pp-wave holographic dictionary 
\be \label{newentry} 
 p_{(m_1, \cdots ,m_k), (n_1, \cdots ,n_k)} = | \bra m_1, \cdots ,m_k | \hat{U} (g) |n_1, \cdots ,n_k\ket |^2, ~~~ k=2,3, 
\ee 
where the operator $\hat{U} (g)$ describes the quantum unitary transition between the degenerate tensionless strings. The BMN single string states form a complete orthonormal basis of the Hilbert space under such finite string interactions 
\be
\sum_{\sum_{i=1}^k n_k=0 }    |n_1,n_2,\cdots ,n_k\ket  \bra n_1,n_2,\cdots ,n_k| =I .
\ee 
 Of course, the probability interpretation only requires the matrix element (\ref{matrixp}) is non-negative. In flat space it is well known that the string coupling constant is related to the vacuum expectation value of a complex axion-dilaton field. In our context of holographic duality, we focus on the correspondence with gauge theory where the effective string coupling $g=\frac{J^2}{N}$ is an arbitrary real non-negative constant. It would seem rather contrived to have negative two-point functions at some higher genus but the total contribution still manages to always remain non-negative. Expecting the same phenomenon as in the case of two string modes, we make the stronger conjecture that the BMN two-point functions with three string modes are also always non-negative separately at each genus. 

If our proposal of the entry of pp-wave holographic dictionary (\ref{newentry}) is correct, to our knowledge, it would not only provide first examples of systematic calculations of (the norms of) the higher genus critical superstring amplitudes, but may also in principle gives exact complete results for any string coupling, due to the convergence of genus expansion, as mentioned in \cite{Huang:2019uue}. Thus it is important to go through the laborious calculations with three string modes at genus two, in order to ensure the previously observed non-negativity at genus one is not just a lucky coincidence, but more likely a manifestation of the deep mathematical structures of the underlying holographic duality.

The paper is organized as the followings. In Sec. \ref{moti} we provide some more discussions on the physical motivation. In Sec. \ref{sectiongenusone}, we review some calculations at genus one. We provide a complete proof of the non-negativity including some cases with mode number degeneracies which require a little extra attentions. In Sec. \ref{sectiongenustwo}, we perform the detailed calculations at genus two, utilizing some symmetries. Although the calculations are quite complicated, we report the result in a relatively compact form in terms of standard integrals, defined in the Appendix \ref{standardint}. We then numerically verify the non-negativity for small modes numbers and give an argument for large mode numbers. We conclude in Sec. \ref{sectionconclusion} with some potential future directions.

\section{More on The Physical Motivation} \label{moti}

Since the paper consists of rather technical calculations, it is useful to explain further about the physical motivation.  A main goal of the theories of quantum gravity is to understand the physics in the regime beyond the reach of classical gravity, e.g. in the highly curved spacetime region near the black hole singularity. For $AdS_d$ space with $d>2$, the scalar curvature is negative $R=-\frac{ d(d-1)}{r^2}$ where $r$ is the radius of the AdS space, so the spacetime is highly curved if the radius $r$ is very small (compared to the string or Planck length). For the pp-wave background (\ref{ppwave}), the scalar curvature actually vanishes, while the non-vanishing component of the Ricci curvature is proportional to $\mu^2$  \cite{Blau:2001ne}.  As we mentioned, we will focus on the $\mu \sim \infty$ or infinite curvature limit. Certainly the effective action of classical gravity completely breaks down in this case, but this is not necessarily a bad situation as we can instead probe the fundamental nature of spacetime. As the background remains maximally supersymmetric, the physics is much more amenable to studies than e.g. those near black hole or big bang singularities. Thanks to the helps from holographic duality, the physical string amplitudes can still be subjected to quantitative studies with some reasonable conjectures and indirect constrains such as unitarity, which is a fundamental principle that we expect to remain valid in the infinite curvature limit. 

In our context, the main goal of the current calculations is to better understand the conjectured entry  (\ref{newentry}) of pp-wave holographic dictionary. This seemingly simple equation has eluded research on the topic for many years, as the physical picture of string dynamics in the infinitely curved pp-wave background turns out to be drastically different from those familiar in flat spacetime or AdS space with large radius. In particular, there is no finite physical process of multiple particles or strings scattering to and from asymptotic region of spacetime, as the higher point functions always vanish in the strict $N\sim \infty$ limit. Instead, the tensionless string directly jumps from one excited state to another through a quantum unitary transition, much as in a S-matrix where the incoming and outgoing states have the same energy. Of course, the BMN higher point functions are still very useful since infinitely many of them may combine to make a finite contribution. The cubic string vertices are in fact the fundamental building blocks of the string (loop) diagrams in the factorization formulas studied in \cite{Huang:2002yt, Huang:2010ne}, but are not needed in the current paper.  Since the only finite BMN two-point functions are always real and symmetric, unitarity of string interactions rules out the naive possibility that they are directly identified with quantum transition amplitudes on the string theory side \cite{Huang:2019lso}. So we arrive at the otherwise seemingly most natural conjecture  (\ref{newentry}). 

There is no technical obstruction for our calculations on the free gauge theory side at any genus, however the available tools on the string theory side are very limited. Much progress for the calculations of higher genus critical superstring amplitudes focused on using the RNS (Ramond-Neveu-Schwarz) formalism in flat space, and is already quite difficult at genus two, see e.g. an early review \cite{DHoker:1988pdl}. Some obstructions to higher genus calculations were discussed more recently in \cite{Donagi:2013dua}. Our conjecture  (\ref{newentry}) gives the norms of certain critical superstring amplitudes including all genus contributions. Of course, the string amplitudes are complex, and consist of the norms and phase angles. As discussed in \cite{Huang:2019lso}, unitarity can in principle determine a large part of the phase angles, but not completely. Our studies thus provide a long term motivation for future research to develop techniques that can deal with string theory on highly curved background, with flux, and including highly excited stringy states, for the purposes of  a direct verification of the conjecture (\ref{newentry}) as well as the complete determination of string amplitudes including the phase angles. Some recent studies concerning highly excited strings are  \cite{Gross:2021gsj, Chen:2021emg}. For the moment, our verifications of non-negativity on the gauge theory side provide indirect non-trivial evidence of the conjecture  (\ref{newentry}). 

We should note that the BMN two-point functions are exactly zero in the cases of the mode numbers $m_i\neq 0, n_i=0$ for an index $i$, as can be easily seen from the integral formula at any genus. This can be understood as a consequence of momentum conservation since a zero mode represents a discretized unit of momentum in the corresponding transverse direction. If we extend the mode numbers to be real, generically there is no indication that these would all be extremal points of the two-point function formula, which may no longer even be real. So the non-negativity seems to be an intrinsically stringy phenomenon, valid only for integer mode numbers satisfying the level matching condition, but can not be extended to real mode numbers. The usual technique of extremization in dealing with functions of complex or real variables is probably not much helpful in our context.

\section{Genus One}  \label{sectiongenusone}

The calculations of the torus BMN two-point functions with many string modes are explained e.g. in our recent paper \cite{Du:2021dml}. The formula is 
\be \ba  \label{freetorus1} 
& \bra \bar{O}^J_{(m_1,m_2,\cdots ,m_k)} O^J_{(n_1,n_2,\cdots ,n_k)} \ket_{\textrm{torus}} \\
&=  g^2 \int_0^1 dx_1dx_2dx_3dx_4 \delta(x_1+x_2+x_3+x_4-1) \int_0^{x_1} dy_k e^{2\pi i (n_k-m_k)y_k} \times \\ 
&  \prod_{i=1}^{k-1}  (\int_0^{x_1} +e^{2\pi i n_i (x_3+x_4)}\int_{x_1}^{x_1+x_2}+e^{2\pi i n_i(x_4-x_2)}\int_{x_1+x_2}^{1-x_4}+e^{-2\pi i n_i(x_2+x_3)}\int_{1-x_4}^1) dy_i e^{2\pi i (n_i-m_i)y_i }
\ea \ee
The result for three string modes were calculated in \cite{Huang:2010ne}, and can be written in terms of the standard integrals in the Appendix \ref{standardint}. For convenience we factor out the coupling constant and denote $  \bra \bar{O}^J_{(m_1,m_2,m_3)} O^J_{(n_1,n_2,n_3)} \ket_{\textrm{torus}}\equiv F_1 g^2 $, with 
\be \ba  \label{F1general}
F_1 &=  \sum_{i\neq j}[  I_{(5,1,1)}(0, m_i-n_i, -m_j+n_j )  +
I_{(2,2,2,1)}(0, -m_j, -m_j+n_j,  m_i-n_i)   \\ & + 
 I_{(2,2,2,1)}(0,  n_j,  -m_j+n_j,  m_i-n_i)    +
 I_{(2,2,1,1,1)}(0, -m_j+n_j  , - m_j,  n_j , m_i-n_i) \\ & +
 I_{(1,1,1,1,1,1,1)}(0, m_i, n_i, -m_j, -n_j , m_i-n_j ,-m_j+n_i )  ]. 
\ea \ee
Since this is a 7-dimensional integral, the indices in the standard integrals always sum to 7. One can explicitly compute the standard integrals using the formulas (\ref{integral3}). In the generic case where none of $m_i, n_i, m_i\pm n_j$'s is zero, there is no further degeneracy in the arguments of the standard integrals in (\ref{F1general}), the result is 
\be  \label{generic1}
F_1 = \frac{\sum_{i=1}^3(m_i-n_i)^2    }{32\pi^4 \prod_{i=1}^3 (m_i-n_i)^2}, 
\ee
which is of course manifestly positive. 

We check also the numerous degenerate cases where some of $m_i, n_i ,m_i\pm n_j$'s vanishes. Most cases also have manifestly non-negative results. However, there are several degenerate cases where the results are somewhat complicated to check the non-negativity, namely the cases 3,7,8 in \cite{Huang:2010ne}. In the followings we perform a more careful analysis to show that they are always positive.  

\begin{enumerate} 
\item $n_3=m_3$ and everything else generic. Using the level matching conditions there are three independent modes numbers. We express the result in terms of $m_1, n_1, m_3$ as 
\be \ba 
F_1   &= \frac{1}{48\pi^2 (m_1-n_1)^2} +\frac{1}{16\pi^4(m_1-n_1)^4}  \\ 
& -\frac{m_3^4+(m_1+n_1 )m_3^3  +m_1n_1 m_3^2 -m_1n_1(m_1+n_1) m_3 -m_1^2 n_1^2 }{16\pi^4 (m_1-n_1)^2 m_1n_1m_3^2(m_3+m_1)(m_3+n_1)}    \\ &
+ \frac{1}{16\pi^4 (m_1-n_1)^2 m_1^2n_1^2m_3^2(m_3+m_1)^2(m_3+n_1)^2}\{  m_3^6 (m_1^2+n_1^2)
 \\ & +2 m_3^5 (m_1^2 n_1+m_1 n_1^2+m_1^3+n_1^3)+m_3^4 (4 m_1^3 n_1+2
   m_1^2 n_1^2+4 m_1 n_1^3+m_1^4+n_1^4)   \\ &
   +2 m_3^3 m_1n_1 (m_1^3+n_1^3)-8  m_3^2m_1^3 n_1^3-4
    m_3 m_1^3 n_1^3 (m_1+n_1)-2 m_1^4 n_1^4
   \}  .
\ea \ee
The expression looks more complicated even than the formula for the generic case (\ref{generic1}). However after some manipulations we can write it as 
\be \ba
F_1   &= \frac{1}{16\pi^4(m_1-n_1)^2 } \{  (\frac{\pi^2}{3} -\frac{1}{m_3^2}) +\frac{1}{(m_1-n_1)^2}   +\frac{m_1^2-m_1n_1+n_1^2}{m_1^2n_1^2} \\ & +\frac{1}{2(m_3+m_1)^2} +\frac{1}{2(m_3+n_1)^2}
   \}  + \frac{1}{32\pi^4(m_3+m_1)^2(m_3+n_1)^2 } .
\ea \ee
Now in this form it is clear that each term is manifestly positive.

\item $n_3=m_3, n_2=m_2$ and everything else generic. The level matching condition also requires $m_1=n_1$. The result is 
\begin{eqnarray}
F_{1} = \frac{1}{120} +\frac{5}{16\pi^4 }(\frac{1}{m_1^4}+\frac{1}{m_2^4}+\frac{1}{m_3^4})-\frac{(m_1^2+m_2^2+m_3^2)^2}{192\pi^2m_1^2m_2^2m_3^2} . 
\end{eqnarray}
There are two independent mode numbers as $m_1+m_2+m_3=0$. Without loss of generality we can assume $m_1, m_2>0$. The negative last term can be estimated 
\be
\frac{(m_1^2+m_2^2+m_3^2)^2}{192\pi^2m_1^2m_2^2m_3^2} = \frac{1}{48\pi^2} [ \frac{m_1^2+m_2^2+m_1m_2}{m_1m_2(m_1+m_2)} ]^2 < \frac{3}{64\pi^2} .
\ee
This is clearly less than the first term, so the result must be overall positive.

\item  $n_3=m_3, n_2=m_1$ and everything else generic. The level matching condition also requires $n_1=m_2$. Correcting a typo in the expression in \cite{Huang:2010ne}, the result is 
\be \ba
F_1 &= \frac{1}{48 \pi ^4 m_1^2 n_1^2
   (m_1-n_1)^4 (m_1+n_1)^2} 
   [9 (m_1^6+n_1^6)  -12m_1n_1(m_1^4+n_1^4)  \\ &
   -9m_1^2n_1^2(m_1^2+n_1^2)  +36m_1^3n_1^3 +\pi^2 m_1^2n_1^2(m_1^2-n_1^2)^2 ] .
   \ea  \ee
It is convenient to denote the positive integers $a=|m_1|, b=|n_1|$. Since $m_1\neq \pm n_1$, we can estimate $(m_1^2-n_1^2)^2= (m_1+n_1)^2  (m_1-n_1)^2 \geq (a+b)^2$. So the numerator is no less than 
\be  \label{estimate2.9}
9 (m_1^6+n_1^6)  -12m_1n_1(m_1^4+n_1^4)  +36m_1^3n_1^3 .
\ee

We can further discuss two cases 
\begin{enumerate}
\item $m_1, n_1$ have the sam sign. Then the expression (\ref{estimate2.9}) is 
\be 
9 (a^6+b^6)  -12ab(a^4+b^4)  
   +36a^3b^3 
  \ee
Using the inequalities 
\be 
2a^6+a^3b^3\geq 3a^5b,   ~~~~ 2b^6+a^3b^3\geq 3ab^5,
\ee
it is easy to see the result is positive. 

\item $m_1, n_1$ have different signs. Then the expression (\ref{estimate2.9}) is 
\be 
9 (a^6+b^6)  +12ab(a^4+b^4)  
   - 36a^3b^3  \geq 6a^3b^3  >0.
  \ee

\end{enumerate} 

\end{enumerate} 

Thus we have provided a complete  proof of non-negativity at genus one for any mode numbers.

\section{Genus Two}  \label{sectiongenustwo}

In this section, we will calculate the genus two BMN two-point functions. At genus $h$, we need to divide the trace in the BMN vacuum operator $\Tr(Z^J)$ into $4h$ parts, and there are $\frac{(4h-1)!!}{2h+1}$ cyclically inequivalent permutations of $(12\cdots (4h))$ representing the Wick contractions of Feynman diagrams \cite{MR848681}. The analysis is quite complicated at genus two. Fortunately, we find some symmetries that will be useful to simplify the calculations. First we will give some explanations about the symmetries, then we use them to write the results in terms of standard integrals.

\subsection{Some Useful Symmetries}

\begin{figure}
\begin{center}
\includegraphics[width=6in]{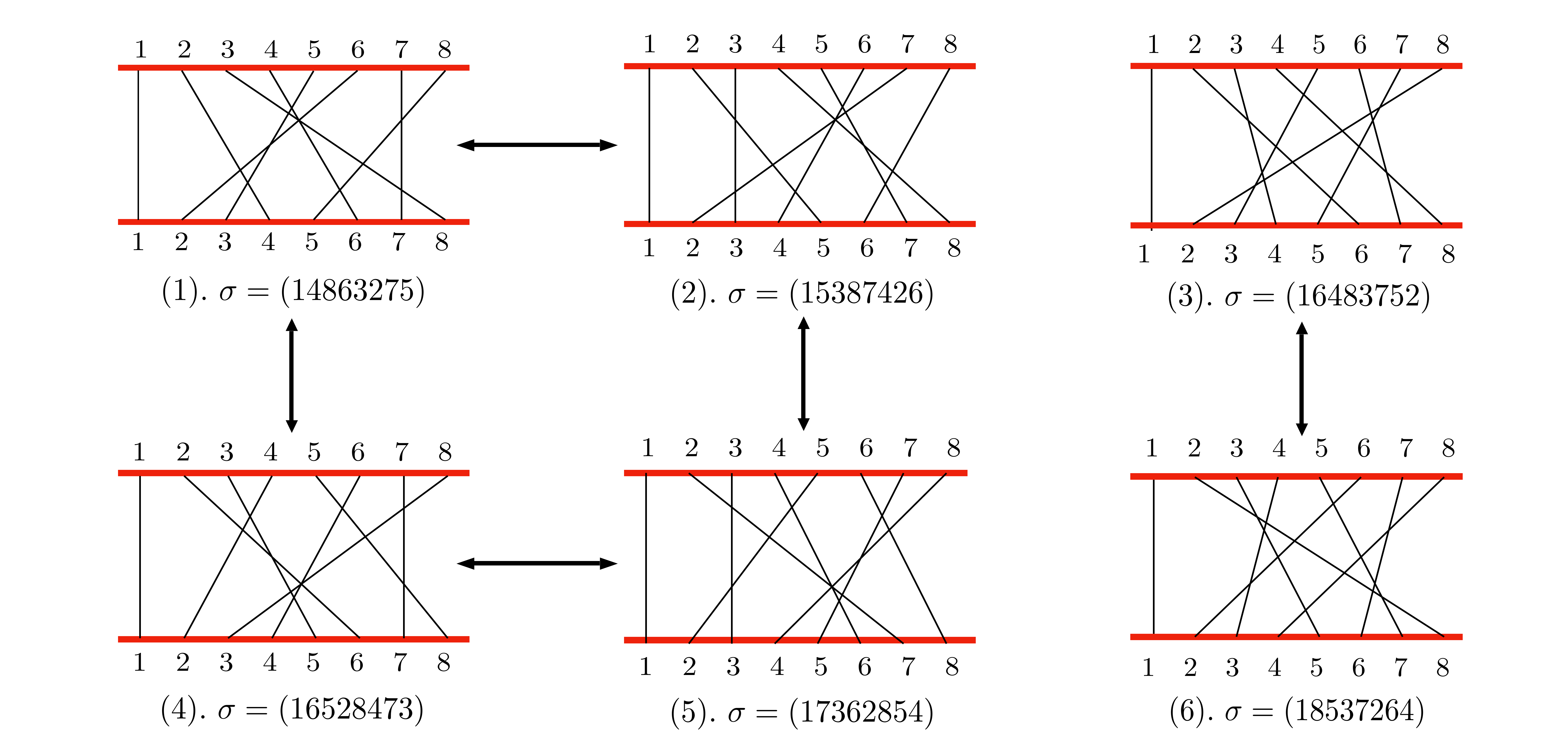} 
\includegraphics[width=6in]{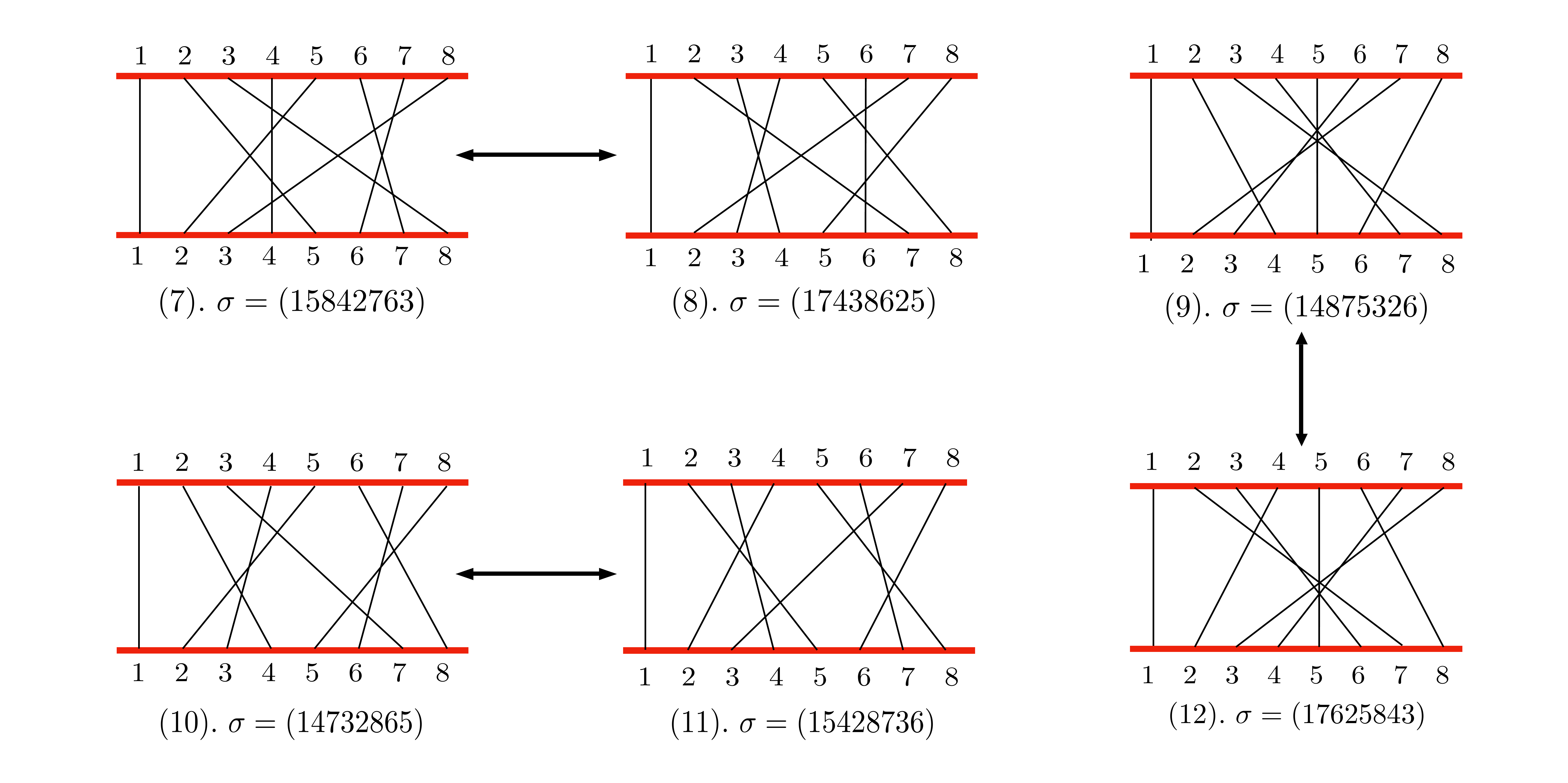}
\includegraphics[width=6in]{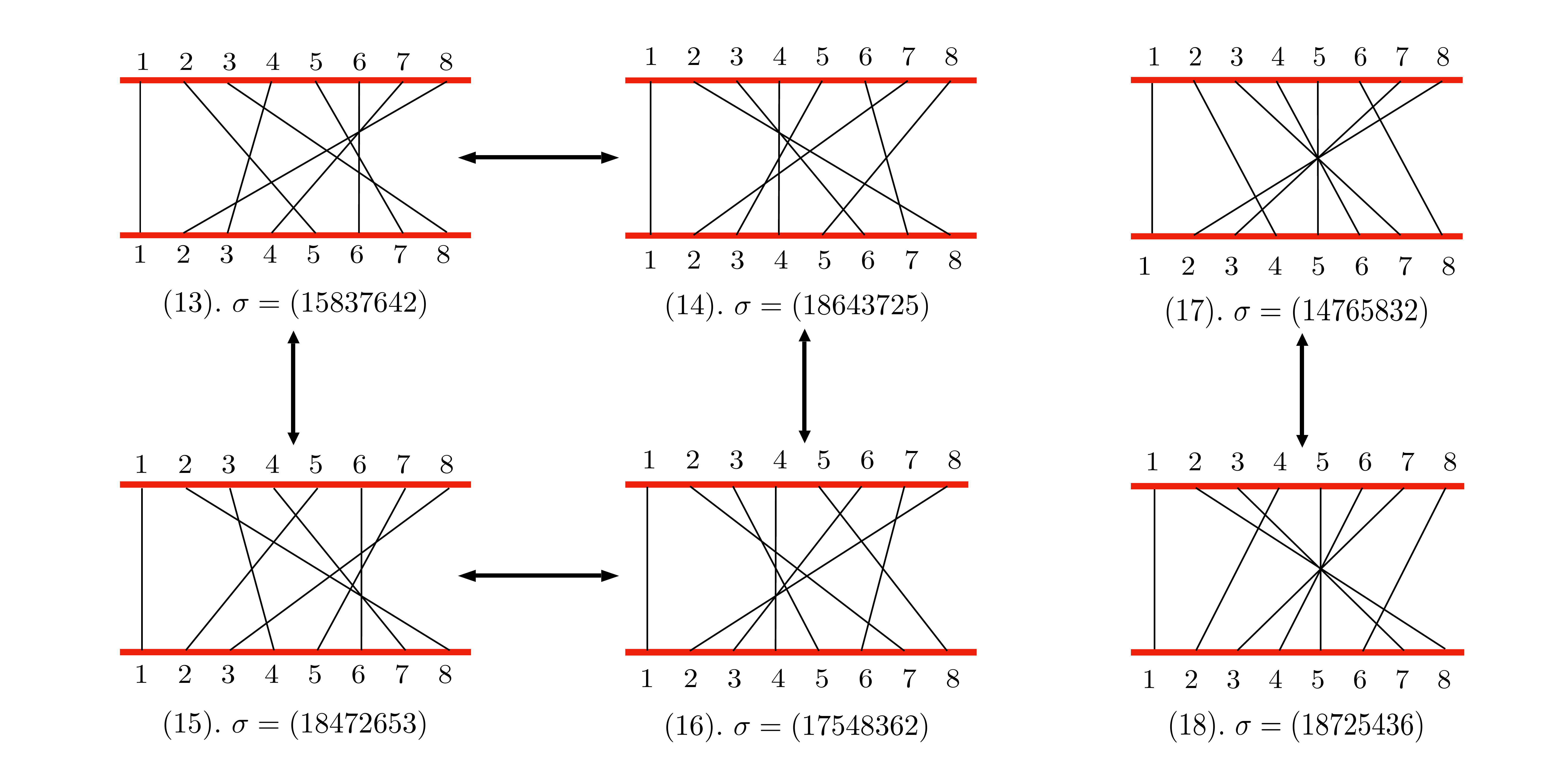}
\end{center}
\end{figure} 

\begin{figure}
\begin{center}
\includegraphics[width=6.5in]{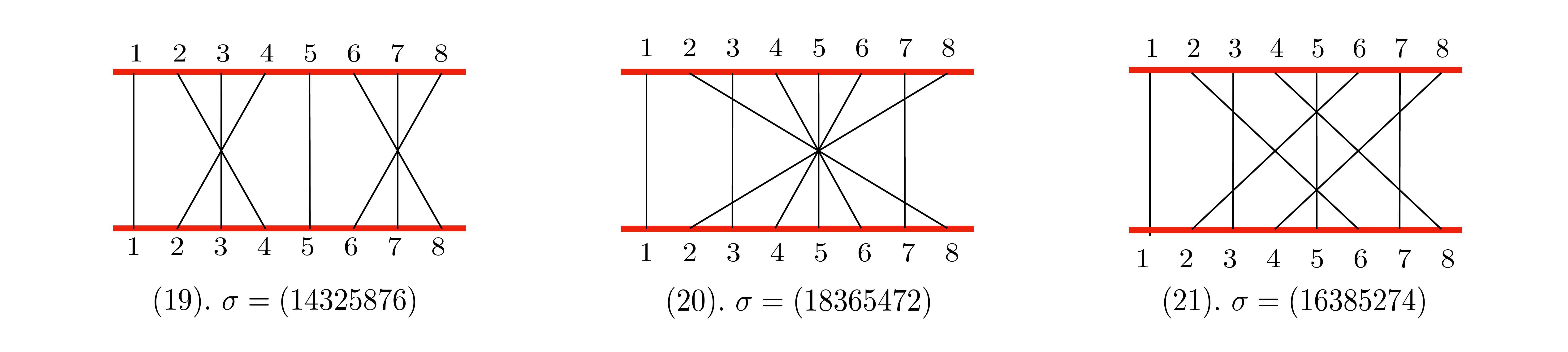}
\end{center}
\caption{The genus 2 diagrams}  \label{genus2}
\end{figure}

We draw the 21 cyclically inequivalent permutations $\sigma_i, i=1,2, \cdots, 21$ in the Fig. \ref{genus2}. For convenience we denote the total genus two contrition in terms of individual permutations as
\be \label{total}
\bra \bar{O}^J_{(m_1,m_2,m_3)} O^J_{(n_1,n_2,n_3)}\ket_{2} = g^4 \sum_{i=1}^{21} F_{\sigma_i} (\vec{m}, \vec{n}) ,
\ee
where $\vec{m} = (m_1,m_2,m_3), \vec{n} =(n_1, n_2, n_3)$.  Here we always fix $\sigma(1)=1$ by cyclic symmetry. We can then insert the three string modes and sum over all positions with phases. We also choose to put the third string mode in the first segment using again cyclic symmetry. The contribution of a particular diagram can be explicitly written as 
\be \ba \label{general}
 F_{\sigma} (\vec{m}, \vec{n}) 
&= \int_{0}^{1} (dx_1 \cdots dx_8)\delta(x_1+\cdots +x_8 -1) \int_{0}^{x_{1}}d y_{3}e^{2\pi i (n_{3}-m_{3})y_{3}}\times \\
& \prod\limits _{k=1}^{2}( \int_{0}^{x_{1}}+ \sum_{j=2}^8
e^{2\pi i n_{k}( \sum\limits_{i=2}^8 s_\sigma(i,j) x_i )}\int_{x_1+\cdots+x_{j-1} }^{x_1+\cdots+x_{j} }
)d y_{k}e^{2\pi i (n_{k}- m_{k}) y_{k}}, 
\ea \ee
where $s_\sigma(i,j)$ is a sign function encoding the phase shifts of the string modes 
\be
s_\sigma(i,j)   =
\begin{cases}
  0  ,      &  i=j, ~~ \textrm{or} ~~ i<j, \sigma(i)<\sigma(j),~~ \textrm{or} ~~ i>j, \sigma(i)>\sigma(j),   \\
  1,       &   i>j, \sigma(i)<\sigma(j),   \\
  -1,      &   i<j, \sigma(i)>\sigma(j). 
\end{cases} 
\ee
For example, the contribution of the first diagram $\sigma= (14863275)$ in Fig. \ref{genus2} is 
\be \ba \label{example1}
F_{\sigma} (\vec{m}, \vec{n}) 
&= \int_{0}^{1} (dx_1 \cdots dx_8)\delta(x_1+\cdots +x_8 -1) \int_{0}^{x_{1}}d y_{3}e^{2\pi i (n_{3}-m_{3})y_{3}}\times \\
& \prod\limits _{i=1}^{2}( \int_{0}^{x_{1}}+e^{2\pi i n_{i}(x_5+x_6)}\int_{x_{1}}^{x_{1}+x_{2}}+e^{-2\pi i n_{i}(x_1+x_2+x_{3})}\int_{x_{1}+x_{2}}^{x_{1}+x_{2}+x_{3}}+\\
			 & e^{2\pi i n_{i}(x_5+x_6+x_8-x_3)}\int_{\sum\limits_{j=1}^3 x_j}^{\sum\limits_{j=1}^4 x_j} +e^{2\pi i n_{i}(x_6-\sum\limits_{j=2}^4 x_j) }\int_{\sum\limits_{j=1}^4 x_j}^{\sum\limits_{j=1}^5 x_j}+e^{-2\pi i n_{i}\sum\limits_{j=2}^5 x_j}\int_{\sum\limits_{j=1}^5 x_j}^{\sum\limits_{j=1}^6 x_j}\\
			 &+e^{2\pi i n_{i}(x_8-x_3) }\int_{\sum\limits_{j=1}^6 x_j}^{1- x_8}+e^{-2\pi i n_{i}(x_3+x_4+x_7) }\int_{1-x_8}^1)d y_{i}e^{2\pi i (n_{i}- m_{i}) y_{i}}. 
		\ea \ee
Note this formalism also works for the simpler genus one case, where the formula (\ref{freetorus1}) comes from the only permutation $(1432)$ in this case. 

A convenient way to check the normalization of (\ref{total}) is to compare with the summation formula (\ref{normalization}). Summing over the $\vec{m}$ string modes can be done by the Poisson summation formula $\sum_{m=-\infty}^{\infty} e^{-2\pi i my} =\sum_{p=-\infty}^{\infty} \delta (y-p)$. The delta functions constrain the non-vanishing contributions to the cases where all string modes are inserted in the same segment. Using the special formula (\ref{special}) it is straightforward to check that for general genus $h$ the sum is indeed $\frac{(4h-1)!!}{2h+1}\frac{1}{(4h)!}= \frac{1}{2^{2h} (2h+1)!} $.

There are two types of involution actions on a permutation $\sigma$, denoted as $\sigma^{-1}$ and $\sigma^\prime$. The $\sigma^{-1}$ is the usual inverse permutation, i.e. $\sigma(\sigma^{-1}(i)) =i$, where $i=1,2,\cdots, 8$, while  $\sigma^\prime$ is a conjugate action discussed in \cite{Du:2021dml}, which maps the permutation $\sigma= (1, a_1, a_2, \cdots, a_7)$ to $\sigma^\prime = (1, 10- a_7, 10- a_6, \cdots, 10-a_1)$.  The two types of actions can be visualized as flipping the diagrams vertically for $\sigma^{-1}$ and horizontally for $\sigma^\prime$ in Fig. \ref{genus2}. Some pairs are related by both actions at the same time, and such relations are not completely explicitly captured in the Fig. \ref{genus2}. Namely we have $\sigma_6=\sigma_3^{-1}=\sigma_3^\prime, \sigma_{11}=\sigma_{10}^{-1}=\sigma_{10}^\prime, \sigma_{18}=\sigma_{17}^{-1}=\sigma_{17}^\prime$. We also have some permutations invariant under the involution actions $\sigma_7=\sigma_7^{-1}, \sigma_8=\sigma_8^{-1}, \sigma_9=\sigma_9^{\prime}, \sigma_{12}=\sigma_{12}^{\prime}, \sigma_{19}=\sigma_{19}^{-1}=\sigma_{19}^{\prime}, \sigma_{20}=\sigma_{20}^{-1}=\sigma_{20}^{\prime}, \sigma_{21}=\sigma_{21}^{-1}=\sigma_{21}^{\prime}$. 

The contributions of the permutations related by the involution actions are related by a simple transformation of the string mode numbers. Specifically, we have 
\be  \label{symme}
F_{\sigma } (\vec{m}, \vec{n}) = F_{\sigma^{-1} } ( -\vec{n}, -\vec{m})  = F_{\sigma^{\prime} } ( -\vec{m}, -\vec{n}). 
\ee
To illustrate we consider the example of the first diagram $\sigma= (14863275)$, which has $\sigma^{-1} = (16528473), \sigma^\prime = (15387426)$. Using the general formula, we can explicitly write 
\be \ba \label{ex1a} 
& F_{\sigma^{-1} } (-\vec{n}, -\vec{m}) \\
&= \int_{0}^{1} (dx_1 \cdots dx_8)\delta(x_1+\cdots +x_8 -1) \int_{0}^{x_{1}}d y_{3}e^{2\pi i (n_{3}-m_{3})y_{3}}\times \\  
& \prod\limits _{i=1}^{2}( \int_{0}^{x_{1}}+e^{ -2\pi i m_{i}(x_3+x_4+x_6+x_8)}\int_{x_{1}}^{x_{1}+x_{2}}+e^{-2\pi i m_{i}(x_4+x_6+x_8-x_2)}\int_{x_{1}+x_{2}}^{x_{1}+x_{2}+x_{3}}+\\
& e^{ 2\pi i m_{i}(x_2+x_3)}\int_{\sum\limits_{j=1}^3 x_j}^{\sum\limits_{j=1}^4 x_j} +e^{- 2\pi i m_{i}\sum\limits_{j=6}^8 x_j }\int_{\sum\limits_{j=1}^4 x_j}^{\sum\limits_{j=1}^5 x_j}+e^{-2\pi i m_{i}(x_8-x_2-x_3-x_5)}\int_{\sum\limits_{j=1}^5 x_j}^{\sum\limits_{j=1}^6 x_j}\\&+e^{-2\pi i m_{i}(x_8-x_5) }\int_{\sum\limits_{j=1}^6 x_j}^{1- x_8}+e^{2\pi i m_{i}(x_1+x_4+x_8) }\int_{1-x_8}^1)d y_{i}e^{2\pi i (n_{i}- m_{i}) y_{i}}.
\ea \ee
We can change the phase factors in (\ref{example1}) from $\vec{n}$ to $\vec{m}$ by a shift of the $y_i$ integration variable. For example we have 
\be \ba
&e^{2\pi i n_{i}(x_5+x_6)}\int_{x_{1}}^{x_{1}+x_{2}}d y_{i}e^{2\pi i (n_{i}- m_{i}) y_{i}}  \\
= ~&e^{2\pi i m_{i}(x_5+x_6)}\int_{x_1+x_5+x_6}^{x_1+x_2+x_5+x_6}d y_{i}e^{2\pi i (n_{i}- m_{i}) y_{i}}. 
\ea \ee
We then rename the integration variables in (\ref{example1}) according to the permutation as 
$x_i  \rightarrow  x_{\sigma(i)}$, or more explicitly $x_1\rightarrow x_1,  x_2\rightarrow x_4, x_3\rightarrow x_8, x_4\rightarrow x_6,
 x_5\rightarrow x_3,  x_6\rightarrow x_2,  x_7\rightarrow x_7,  x_8\rightarrow x_5$. It is easy to check that the expressions (\ref{example1})  and (\ref{ex1a}) are indeed equal. 

The case of $\sigma^\prime$ was discussed in \cite{Du:2021dml}, and we give some more details here. For the example we can write 
\be \ba \label{ex1b} 
&F_{\sigma^{\prime} } (-\vec{m}, -\vec{n}) \\
&= \int_{0}^{1} (dx_1 \cdots dx_8)\delta(x_1+\cdots +x_8 -1) \int_{0}^{x_{1}}d y_{3}e^{- 2\pi i (n_{3}-m_{3})y_{3}}\times \\  
& \prod\limits _{i=1}^{2}( \int_{0}^{x_{1}}+e^{ -2\pi i n_{i}(x_3+ x_6+x_7)}\int_{x_{1}}^{x_{1}+x_{2}}+e^{-2\pi i n_{i}(x_7 -x_2)}\int_{x_{1}+x_{2}}^{x_{1}+x_{2}+x_{3}}+\\
& e^{ -2\pi i n_{i}(\sum\limits_{i=5}^8 x_i )}\int_{\sum\limits_{j=1}^3 x_j}^{\sum\limits_{j=1}^4 x_j} +e^{- 2\pi i n_{i}(x_6+x_7+x_8-x_4) }\int_{\sum\limits_{j=1}^4 x_j}^{\sum\limits_{j=1}^5 x_j}+e^{-2\pi i n_{i}(x_7-x_2-x_4-x_5)}\int_{\sum\limits_{j=1}^5 x_j}^{\sum\limits_{j=1}^6 x_j}\\&+e^{ 2\pi i n_{i}(\sum\limits_{j=2}^6 x_j) }\int_{\sum\limits_{j=1}^6 x_j}^{1- x_8}+e^{2\pi i n_{i}(x_4+x_5) }\int_{1-x_8}^1)d y_{i}e^{- 2\pi i (n_{i}- m_{i}) y_{i}}.
\ea \ee
We change the integration variables in (\ref{example1}) by $y_i\rightarrow 1+x_1-y_i, ~i=1,2,3$. This gives an overall phase which is trivial by closed string level matching conditions $\exp[ 2\pi i \sum\limits_{i=1}^3(n_i-m_i)x_1]=1$. The first integration range $[0,x_1]$ is not changed since the $y_i$ can be always shifted by an integer. The other 7 integrals are reversed in order as visualized by horizontally flipping the diagram. We further change the integration variables $x_i\rightarrow x_{10-i}, ~i=2,3,\cdots 8$ with $x_1$ unchanged. It is not difficult to check that the integrals (\ref{example1}) and (\ref{ex1b}) are also equal. 

It is clear that the $\vec{m}\rightarrow -\vec{m}, \vec{n}\rightarrow -\vec{n}$ transformation is the complex conjugate action.  Using together with the symmetry of inverse permutation (\ref{symme}),  since the set of 21 diagrams in Fig. \ref{genus2} is closed under both involution actions, we have explicitly verified the two-point functions are real and symmetric at genus two. Similarly this works at any higher genus $h$ since flipping the diagrams vertically or horizontally does not change the genus
\be \ba
&\bra \bar{O}^J_{(m_1,m_2,m_3)} O^J_{(n_1,n_2,n_3)}\ket_{h}^*=\bra \bar{O}^J_{(m_1,m_2,m_3)} O^J_{(n_1,n_2,n_3)}\ket_{h} \\
&\bra \bar{O}^J_{(m_1,m_2,m_3)} O^J_{(n_1,n_2,n_3)}\ket_{h}=\bra \bar{O}^J_{(n_1, n_2, n_3)} O^J_{(m_1,m_2,m_3)}\ket_{h}.
\ea \ee

\begin{figure}
\begin{center}
\includegraphics[width=6.5in]{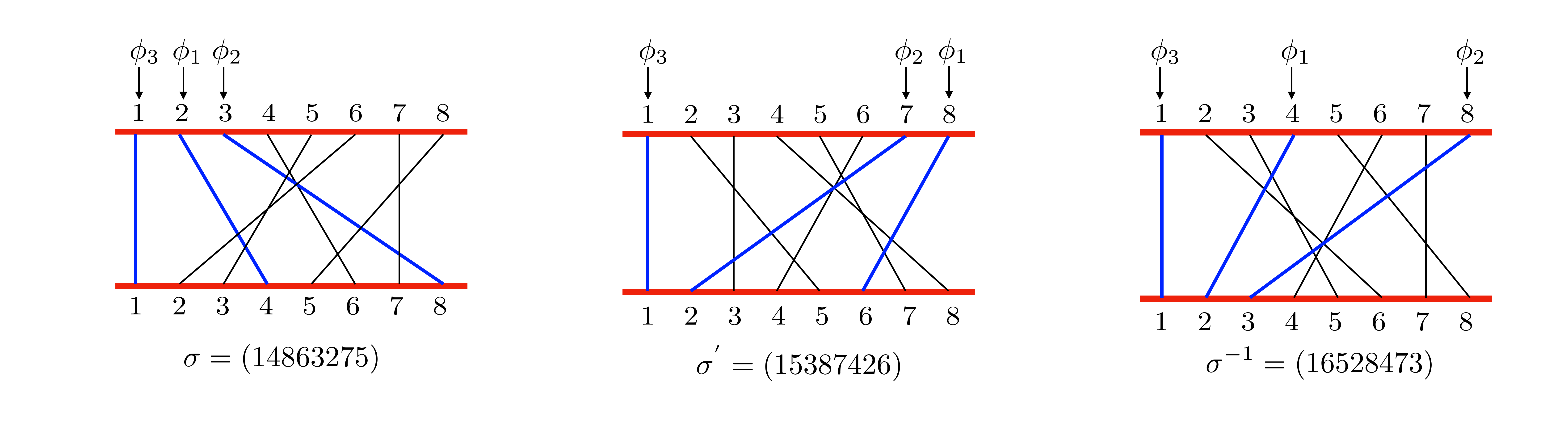} 
\end{center}
\caption{The symmetry relations with three string insertions}  \label{mnsymmetry}
\end{figure}

In fact, there are some more refined symmetries with specified positions for string modes. First we introduce a notation $F_\sigma(\vec{m}, \vec{n}) (i, j, k)$, denoting the contribution of a permutation $\sigma$ with the three string modes inserted in the $i,j,k$'th segments. The formula (\ref{general}) is then simply $F_\sigma(\vec{m}, \vec{n}) = \sum_{i,j=1}^8 F_\sigma(\vec{m}, \vec{n})(i,j, 1)$. Suppose $P$ is one of 6 permutations of $(123)$, then we note that the contribution is obviously invariant under the permutation of the three positions $(i,j,k)$ and mode numbers simultaneously, namely $F_\sigma(P(\vec{m}), P(\vec{n})) (P(i, j, k)) = F_\sigma(\vec{m}, \vec{n}) (i, j, k)$. 

Similarly as the symmetry formulas (\ref{symme}), we can derive the more refined formulas 
\be\ba
F_\sigma(\vec{m}, \vec{n}) (i,j,k) &=F_{\sigma^{'}}(-\vec{m},-\vec{n}) (10-i,10-j,10-k) \\
&= F_{\sigma^{-1}}(-\vec{n},-\vec{m}) (\sigma(i),\sigma(j),\sigma(k)), 
\ea\ee
where the three positions are understood to be equivalent mod 8.  As an example, in Figure (\ref{mnsymmetry}), 
$\sigma=(14863275)$, three string modes $(\phi_1,\phi_2,\phi_3)$ is inserted into the segments $(2,3,1)$, and we have the symmetry relation 
\be\ba
F_\sigma(\vec{m}, \vec{n})(2,3,1) =F_{\sigma^{'}}(-\vec{m},-\vec{n}) (8,7,1) =F_{\sigma^{-1}}(-\vec{n},-\vec{m}) (4,8,1). 
\ea\ee
So from one contribution, we can infer the results of the other two. In this case their formulas in terms of standard integrals are  
\be\ba
 F_\sigma(\vec{m},\vec{n})(2,3,1) &=I_{(3,2,2,2,2)}(0,n_1,-n_2,-m_2,n_1+m_3), \\
F_{\sigma^{'}}(\vec{m},\vec{n}) (8,7,1)  &=I_{(3,2,2,2,2)}(0,-n_1,n_2,m_2,-n_1-m_3), \\
F_{\sigma^{-1}}(\vec{m},\vec{n}) (4,8,1) &=I_{(3,2,2,2,2)}(0,-m_1,m_2,n_2,-n_3-m_1). 
\ea\ee

\begin{figure}
\begin{center}
\includegraphics[width=6.5in]{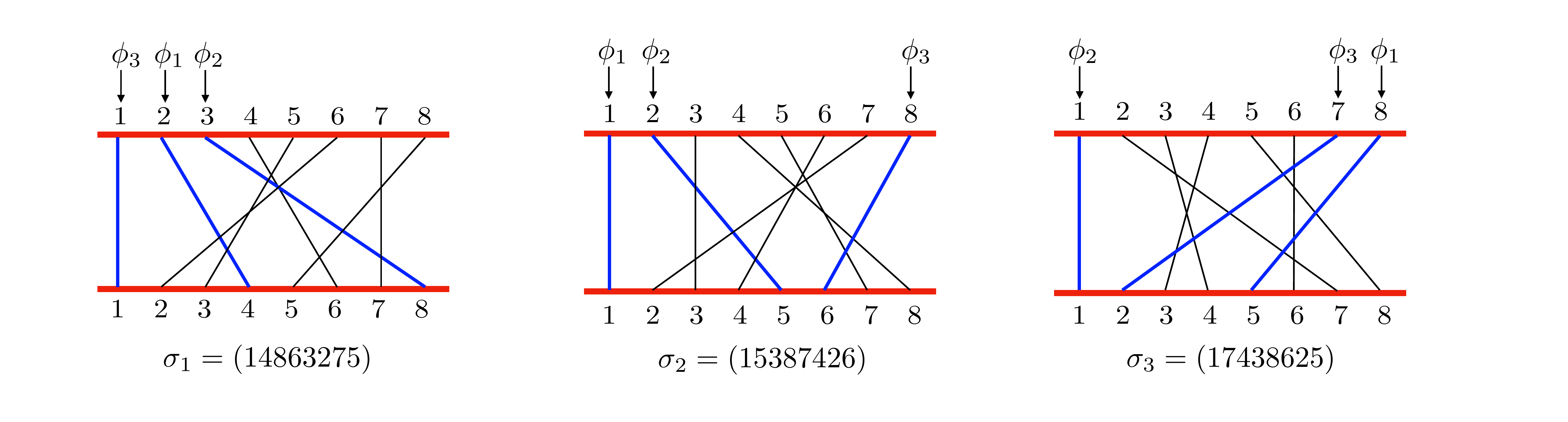} 
\end{center}
\caption{The cyclical symmetry }  \label{cyclicalsymmetry}
\end{figure}

Finally, there is a useful residual cyclical symmetry.  When writing two-point function, we have fixed the third string mode into the first  segment using the cyclicity, which breaks the symmetry between different modes. But due to the permutation symmetry of string modes, there is still some useful relations for calculations.  For a permutation $\sigma$, if we insert $(\phi_1,\phi_2,\phi_3)$ into the positions $(i,j,1)$, and rotate the 8 segments  such that the $\sigma(i)$ or $\sigma(j)$ is put into the first segment, the integral contribution is still the same, so we have 
\be\ba
F_\sigma(\vec{m},\vec{n}) (i,j,1) &=F_{\sigma-\sigma(i)+1}(\vec{m},\vec{n}) (1,j-i+1,2-i) \\
&=F_{\sigma-\sigma(j)+1} (\vec{m},\vec{n} ) (i+1-j,1,2-j) 
\ea\ee
where $\sigma-i$ means that we subtract $i$ from each number of the permutation $\sigma$, and using the cyclicality to put 1 into the first place. Again it is implicit that the numbers are equivalent mod 8. 

As an example, consider the three permutations in Fig. \ref{cyclicalsymmetry}, where $\sigma_1=(14863275)$ and  $\sigma_2 = \sigma_1-3, \sigma_3 =\sigma_1+1$.  We have the following relations also with results in terms of a standard integral
\be\ba
F_{\sigma_1}( \vec{m},\vec{n} ) (2,3,1) =&I_{(3,2,2,2,2)}(0,n_1,-n_2,-m_2,n_1+m_3)\\
=&F_{\sigma_2} ( \vec{m},\vec{n} ) (1,2,8)\\
=&F_{\sigma_3} ( \vec{m},\vec{n} ) (8,1,7) . 
\ea\ee
We see this symmetry relates the different diagrams in the group with the same string diagram multiplicity in the classification in \cite{Du:2021dml}. In our calculations we always put $\phi_3$ into the first segment in every diagram, we can use the above formulas to nicely combine the contributions into sums over the permutations of the three string modes. For the current example we have 
\be\ba
&F_{\sigma_1}(\vec{m},\vec{n})(2,3,1) + F_{\sigma_1}(\vec{m},\vec{n}) (3,2,1) + F_{\sigma_2}(\vec{m},\vec{n}) (2,8,1) \\
+&F_{\sigma_2}(\vec{m},\vec{n} ) (8,2,1) + F_{\sigma_3} (\vec{m},\vec{n}) (8,7,1) + F_{\sigma_3} (\vec{m},\vec{n}) (7,8,1) \\
=&\sum\limits_{(i,j,k)}I_{(3,2,2,2,2)}(0,n_i,-n_j,-m_j,n_i+m_k), 
\ea\ee 
where the sum $(i,j,k)$ is over the 6 permutations of (123).

\subsection{The Result}

We can classify the 21 diagrams in Fig. (\ref{genus2}) into four groups according to \cite{Du:2021dml}, then we can use the two involution symmetries to further separate the diagrams into small groups as  
\be \ba \label{biggroups}
&8:\{\{(1),(2),(4),(5)\},~~\{(3),(6)\},~~\{(7),(8)\}\}\\
&8:\{\{(9),(12)\},~~\{(10),(11)\},~~\{(13),(14),(15),(16)\}\}\\
&4:\{\{ (17),(18)\},~~\{(19)\},~~\{(20)\}\}\\
&1:\{(21)\}.
\ea\ee
To calculate the total contribution (\ref{total}), we only need to choose one permutation in each small group, e.g.  
	 $$(1),(3),(7),(9),(10),(13),(17),(19),(20),(21). $$
The contributions of other diagrams can then be obtained by symmetry. Each permutation consists of $\frac{8\cdot 9}{2}= 36$ different standard integrals, so the computation is still very complicated. 

After some very lengthy calculations, we finally obtain the result in terms of standard integrals, which is organized into four parts as 
{\footnotesize 
\be\ba \label{I1}
& ~~ I_1( \vec{m},\vec{n} ) \\  & = \sum\limits_{(i,j,k)} 21 I_{(9,1,1)}(0,n_i-m_i,-n_j+m_j) 
 +  I_{(5, 5, 1)} (0, n_i - m_i, -n_j + m_j)    \\
&+  I_{(5, 2, 2, 1, 1)}  (0, n_i - m_i, -n_j + m_j,  n_i + m_j, -n_j - m_i )  \\
&+  I_{(5, 1, 1, 1, 1, 1, 1)} (0, n_i - m_k, -n_k + m_i, m_i,   n_i, -m_k, -n_k) \\
&+  6  I_{(3, 3, 2, 2, 1)} (0, n_i - m_i,   n_i, -m_i, -n_j + m_j )    \\
&+  2 I_{(3, 2, 2, 1, 1, 1, 1)} (0, n_j - m_j, -n_k + m_k,   n_j, -n_k, -m_j, m_k)    \\
&+  I_{(2, 2, 1, 1, 1, 1, 1, 1, 1)} (0, -n_k + m_k, -n_k, m_k, n_i, -m_j, -n_k - m_j,    n_i + m_k, n_i - m_j )  ,  
\ea\ee
\be\ba
& ~~ I_2( \vec{m},\vec{n} ) \\  & = \sum\limits_{(i,j,k)}  5 I_{(6, 2, 2, 1)} (0, -m_i, -n_i, n_k + m_j) 
 +  5  I_{(6, 2, 1, 1, 1)} (0, -n_i + m_i, -n_i, m_i,   n_k - m_k)    \\
&+  I_{(5, 2, 2, 2)} (0, n_i + m_j, n_i - m_i, -n_j + m_j )   
 +  I_{(5, 2, 2, 1, 1)} (0, m_i, n_i, n_i + m_i, -n_j - m_k )   \\
&+  I_{(5, 1, 1, 1, 1, 1, 1)}  (0, -n_k, m_j, -n_k + m_j,   n_i + m_j, n_i - m_i, -n_k - m_i)    \\
&+  2  I_{(4, 3, 3, 1)} (0, -m_i, -n_i, n_k + m_j )  
\\& + 3  I_{(4, 2, 2, 2, 1)} (0, m_i, -n_i, -n_i + m_i,   n_k - m_k)    \\
&+  I_{(4, 1, 1, 1, 1, 1, 1, 1)}  (0, -m_k, n_k, -n_j - m_k,   n_k + m_i, m_i, -n_j, n_k - m_k )     \\
&+  4   I_{(3, 3, 3, 1, 1)} (0, n_i - m_i, -m_i,   n_i, -n_j + m_j )  \\
&+  2   I_{(3, 2, 2, 2, 2)} (0, -n_k, -m_j,   n_j - m_j, -n_k + m_k )  \\
&+ 2 I_{(3, 2, 2, 2, 1, 1)} (0, n_k,   m_k, -n_i - m_j, -n_i, -m_j ) \\
&+  I_{(3, 2, 2, 1, 1, 1, 1) } (0, n_i, -m_j, -n_k - m_j,   n_i - m_j, n_i + m_k, -n_k + m_k )  \\
&+  I_{(3, 2, 2, 1, 1, 1, 1)}  (0, n_j, -m_i, -n_i, m_j,   n_j - m_i, -n_i + m_j )   \\
&+  I_{(3, 2, 2, 1, 1, 1, 1)}   (0, n_j, m_j, -n_i, -m_i,   n_j - m_i, -n_i + m_j )  \\
&+  I_{(2, 2, 2, 2, 2, 1)}  (n_j + m_i, m_i, -n_k, n_j, -m_k,   0)    \\
&+  I_{(2, 2, 2, 2, 1, 1, 1)} (0, n_j - m_j, -n_k + m_k,   m_k, -n_k, -m_j, n_j + m_k )  \\
&+  I_{(2, 2, 2, 2, 1, 1, 1)} (0, -n_k, -m_j, -n_k - m_j, n_i,   n_i + m_k, -n_k + m_k )      \\
&+ 2  I_{(2, 2, 2, 1, 1, 1, 1, 1) } (0,   n_j - m_j, -n_k + m_k, -n_k, n_j, -m_j, -n_k - m_j,   n_j + m_k )    \\
&+  I_{(2, 2, 2, 1, 1, 1, 1, 1)} (0, -n_k, -n_k + m_k,   n_i, -m_j, -n_k - m_j, n_i - m_j, n_i + m_k )      \\
&+ 2 I_{(2, 2, 2, 1, 1, 1, 1, 1)} (0, -n_k, -n_k + m_k, n_i, -n_k - m_j, -m_j, m_k,    n_i + m_k ), 
\ea\ee

\be\ba
& ~~ I_3( \vec{m},\vec{n} ) \\  & = \sum\limits_{(i,j,k)}   
I_{(3, 3, 1, 1, 1, 1, 1)} (0, n_i, -n_k, -m_j, -n_k - m_j,   n_i + m_k, -n_k + m_k ) \\
&+ 2  I_{(3, 2, 2, 2, 2)} (0, n_i, -n_j,   n_i - m_i, -n_j + m_j )  \\
&+  I_{(3, 2, 2, 1, 1, 1, 1)} (0, m_j, -m_i,   n_j, -n_i, -m_i + n_j, m_j - n_i )   \\
&+  I_{(2, 2, 1, 1, 1, 1, 1, 1, 1)} (0, m_k, n_j,   n_j - m_j, -m_j, -n_i - m_j, -n_i + m_k, n_j + m_k, -n_i ) ,   \\
\ea\ee	 	
 
\be\ba
& ~~ I_4( \vec{m},\vec{n} ) \\  & = \sum\limits_{(i,j,k)}   
5 I_{(6, 2, 2, 1)} (0, -m_i, n_i - m_i, -n_j + m_j )   
+ 4   I_{(5, 3, 2, 1)} (0, m_i, -n_i + m_i, n_k - m_k)  \\
&+ 4 I_{ (5, 3, 2, 1)} (0, n_i, m_i, -n_j - m_k )    
+ 4 I_{ (5, 2, 2, 1, 1)} (0, n_i,   n_i - m_i, -m_i, -n_j + m_j )     \\
&+  I_{ (5, 1, 1, 1, 1, 1, 1)} (0, -n_k, n_i, m_j, n_i + m_j,   n_i - m_i, -n_k - m_i )  \\
&+  3  I_ {(4, 4, 2, 1)} (0, n_i, n_i - m_i, -n_j + m_j ) 
+  3  I_{(4, 3, 2, 1, 1)} (0, n_i,   n_i - m_i, -m_i, -n_j + m_j )   \\
&+ 2  I_{(4, 3, 2, 1, 1)} (0, -m_i, -n_i, -n_i - m_i,   n_k + m_j )    \\
&+  I_{(4, 2, 1, 1, 1, 1, 1)} (0, m_i, -m_k, -n_j, -m_k - n_j,   m_i + n_k, n_k - m_k )     \\
&+  I_{(4, 2, 1, 1, 1, 1, 1)} (0, n_j, m_k, -n_i,   n_j - m_j, -n_i - m_j, n_j + m_k )   \\
&+  I_{(4, 2, 1, 1, 1, 1, 1)} (0, -m_j, -n_j, m_k, n_k,   n_k - m_j, -n_j + m_k )  \\
&+  I_{(4, 2, 1, 1, 1, 1, 1)} (0, -n_k, -m_i, m_j, n_i + m_j,   n_i - m_i, -n_k - m_i )    \\
&+  I_{(4, 2, 1, 1, 1, 1, 1)} (0, m_k, -n_i,   n_j - m_j, -n_i - m_j, -n_i + m_k, n_j + m_k )    \\
&+  I_{(3, 3, 1, 1, 1, 1, 1)} (0, -n_k, n_i, -m_j, -n_k - m_j,   n_i + m_k, -n_k + m_k )     \\
&+  I_{(3, 3, 1, 1, 1, 1, 1)} (0, n_i, -m_j, -n_k - m_j,   n_i - m_j, n_i + m_k, -n_k + m_k )  \\
&+ 2   I_{(3, 2, 2, 2, 2)} (0, n_i, -m_j, -n_j, n_i + m_k)    \\
&+ 2    I_{(3, 2, 2, 2, 1, 1)} (0, n_j - m_j, -n_k + m_k, m_k,   n_j, -m_j )  \\
&+  I_{(3, 2, 2, 1, 1, 1, 1)}  (0, n_i, -n_k, n_i + m_j,   n_i - m_i, -n_k - m_i, m_j )    \\
&+  I_{(3, 2, 2, 1, 1, 1, 1)}  (0, m_i, n_j, -m_k,   n_j + m_i, -n_i + m_i, -n_i - m_k )    \\
&+  I_{(3, 2, 2, 1, 1, 1, 1)}  (0, n_i, -m_j, -n_k - m_j,   n_i + m_k, m_k, -n_k + m_k )    \\
&+  I_{(3, 2, 1, 1, 1, 1, 1, 1)} (0, -n_k, n_i + m_j, n_i,   n_i - m_i, -m_i, -n_k - m_i, m_j )   \\
&+  I_{(3, 2, 1, 1, 1, 1, 1, 1)} (0, m_j, -n_j + m_j, -n_j,   n_k, -n_j - m_i, n_k + m_j, -m_i )    \\
&+  I_{(2, 2, 2, 2, 2, 1)} (0, -n_k,   n_j - m_j, -n_k - m_j, -n_k + m_k, n_j )    \\
&+  I_{(2, 2, 2, 2, 2, 1)} (0, n_j, -n_k,   n_j - m_j, -n_k + m_k, -m_j )   \\
&+  2  I_{(2, 2, 2, 2, 1, 1, 1)} (0,   n_j - m_j, -m_j, -n_k + m_k, -n_k, -n_k - m_j, m_k )   \\
&+  I_{(2, 2, 2, 1, 1, 1, 1, 1)} (0, -n_k - m_j, -m_j, -n_k,   n_i, n_i - m_j, n_i + m_k, -n_k + m_k )   \\
&+  I_{(2, 2, 2, 1, 1, 1, 1, 1)} (0, -m_j, m_k, -n_j, n_k,   n_k - m_j, -n_j + m_k, m_k + n_k )  \\
&+  I_{(2, 2, 2, 1, 1, 1, 1, 1)} (0, n_j, -m_i,   m_j, -n_i, -n_i - m_i, n_j - m_i, -n_i + m_j ),    
\ea\ee 
}
where $(i,j,k)$ sums over the 6 permutations of (123), so each term is apparently invariant under permutations of the three string modes. There are  7, 20, 4, 28 elements in $I_1, I_2, I_3, I_4$ with a total of 59 elements. The total contribution to the two-point function includes the elements in $I_1, I_2, I_3, I_4$ and their transformations as  
\be  \ba \label{totalresult} 
& ~~~~ g^{-4} \bra \bar{O}^J_{(m_1,m_2,m_3)} O^J_{(n_1,n_2,n_3)}\ket_{2} \\
& = I_1( \vec{m},\vec{n} ) + I_2( \vec{m},\vec{n} ) 
+ I_2(- \vec{m}, - \vec{n} ) +I_3( \vec{m},\vec{n} ) + I_3( -\vec{n}, -\vec{m} ) \\ & + I_4( \vec{m},\vec{n} )  + I_4( -\vec{m}, -\vec{n} ) +
I_4( -\vec{n}, -\vec{m} ) +I_4( \vec{n}, \vec{m} )  \\
&= I_1( \vec{m},\vec{n} ) + I_3( \vec{m},\vec{n} ) + I_3( -\vec{n}, -\vec{m} ) +2\Re [I_2( \vec{m},\vec{n} )+ I_4( \vec{m},\vec{n} ) + I_4( \vec{n},\vec{m} ) ]. 
\ea \ee
So together with their transformations, each element in $I_1, I_2, I_3, I_4$ contributes 1,2,2,4 times. In the second equality we have used the fact that the transformation $(\vec{m}, \vec{n}) \rightarrow -(\vec{m}, \vec{n})$ is the complex conjugation. 
	
We can analyze the reality property. Each element in $I_1( \vec{m},\vec{n} ) , I_3( \vec{m},\vec{n} ), I_3( -\vec{n}, -\vec{m} )$ summing over the $(i,j,k)$ permutations can be easily shown to be real using some obvious symmetries of the standard integrals. On the other hand,  each element in $I_2( \vec{m},\vec{n} ) , I_4( \vec{m},\vec{n} ), I_4( \vec{n}, \vec{m} )$ summing over the $(i,j,k)$ permutations is not always real with the exception of one element in $I_2$, namely we can explicitly check that the reality of the following sum
\be
  \sum\limits_{(i,j,k)} I_{(2, 2, 2, 1, 1, 1, 1, 1) } (0,   n_j - m_j, -n_k + m_k, -n_k, n_j, -m_j, -n_k - m_j,   n_j + m_k ),  
  \ee	
though this is not easily derived by obvious symmetries of the standard integrals. 

For the generic case where none of $m_i, n_i ,m_i\pm n_j$ is zero, we note that the element with indices at most 2, i.e. the standard integral $I_{(2,\cdots )}$ appearing here, always has vanishing real part. This is not obvious from symmetries of the standard integrals but can be checked explicitly. This property is no longer always true if there are some mode number degeneracies.

We can perform some tests on the rather complicated final result (\ref{totalresult}). For example, when two string modes are zero $m_3=n_3=0$, the result correctly reduces to the case of two string modes, previously first obtained  in  \cite{Constable:2002hw}. For the case $m_3=0, n_3\neq  0$, the result correctly vanishes. We can also test numerically the summation formula (\ref{normalization}). We fix one set of string modes and perform the sum over the other set of modes numerically. As an example, we check that the two-point functions with $\vec{m}=(1,1,-2)$ and all $\vec{n}$ modes in the range $\max(|n_i|)\leq1424$ are always non-negative, and their numerical sum is 0.999001 of the exact value $\frac{1}{1920}$.

It seems far from obvious the expression (\ref{totalresult}) is always non-negative. We check numerically for all cases with mode numbers $\max(|m_i|, |n_i|)\leq 30$. Furthermore, we also test some random mode numbers in larger range. For example, we take random integers $m_i, n_i, i=1,2$ from a Gaussian distribution with expectation value 0 and standard deviation 100, while $m_3=-m_1-m_2, n_3 =-n_1-n_2$ are then determined. We compute the result (\ref{totalresult}) for more than a million random sets of mode numbers in this way.  In all tests we have not found any negative result.

For the generic case, an analytic expression can be obtained, but it is too long to write down completely for analysis. We find that the result (\ref{totalresult}) can be written as 
\be 
f_{-4} (\vec{m}, \vec{n}) +f_{-6} (\vec{m}, \vec{n}) +f_{-8} (\vec{m}, \vec{n})  ,
\ee 
where $f_k(\vec{m}, \vec{n})$ is a rational function of the string mode numbers with homogeneous degree $k$, which is also the degree of $\pi$, the only irrational number appearing in the final result. The expressions $f_{-6}, f_{-8}$ are quite complicated, and can be either positive or negative. However, we find that the $f_{-4}$ expression is quite simple and is actually related to the genus one result (\ref{generic1}) up to a numerical constant 
\be 
f_{-4} (\vec{m}, \vec{n}) = \frac{7 \sum_{i=1}^3(m_i-n_i)^2    }{3840\pi^4 \prod_{i=1}^3 (m_i-n_i)^2}. 
\ee
We also note that $f_{-4}$ comes from the element with most degenerate indices, namely the first element $I_{(9,1,1)}(0,n_i-m_i,-n_j+m_j) $ in $I_1$ in (\ref{I1}), which in turns comes from the contributions with all three string modes in the same of the 8 segments. For string mode numbers with large absolute values, the manifestly positive term $f_{-4} (\vec{m}, \vec{n})$ should be dominant, so it is indeed plausible that the total result should remain positive in this case. 

We expect this argument works also at general genus $h$. For generic mode numbers with large absolute values, the dominant contributions should come from the cases in the $\frac{(4h-1)!!}{2h+1}$  permutations where all string modes are in the same segment. For generic case we can first perform the $dy$ integrals in e.g. (\ref{general}) for the genus two case. Using the formulas in Appendix \ref{standardint}, it is not difficult to calculate the leading term 
\be  \ba
& ~~~\bra \bar{O}^J_{(m_1,m_2,m_3)} O^J_{(n_1,n_2,n_3)}\ket_{h} \\ &= g^{2h} \frac{(4h-1)!!}{2h+1} 
 \int_{0}^{1} (dx_1 \cdots dx_{4h})\delta(x_1+\cdots +x_{4h} -1) \prod_{i=1}^3 \frac{e^{2\pi i (n_i-m_i)x_1}-1}{2\pi i(n_i-m_i)} +\cdots \\
 &  =\frac{g^{2h}}{(2\pi i)^3} \frac{(4h-1)!!}{2h+1}  (\prod_{i=1}^3 \frac{1}{n_i-m_i }) \sum_{i=1}^3 [I_{(4h-1,1)}(0, n_i- m_i) - I_{(4h-1,1)}(0, m_i-n_i) ]+\cdots  \\
& = g^{2h} \frac{h(4h-1)}{2^{2h+2}(2h+1)!}  \frac{ \sum_{i=1}^3(m_i-n_i)^2    }{\pi^4 \prod_{i=1}^3 (m_i-n_i)^2} +\cdots ,
\ea \ee
which is also proportional to the genus one formula (\ref{generic1}) and manifestly positive. Of course, this argument does not apply to the cases with small absolute values of string modes, where the non-negativity seems highly non-trivial. 

The similar phenomenon actually also appears in the calculations in the case of two and four string modes in previous papers. The dominant contribution for generic mode numbers with large absolute values also comes from the situations where all string modes are inserted in the same segment. In these cases of an even number $k$ of string modes, the dominant contribution has the same sign as $(2\pi i)^k \prod_{i=1}^k (n_i-m_i )$, which is always positive for $k=2$, but can be either positive or negative for $k\geq 4$.

For the case of two string modes, the non-negativity is valid separately for groups of diagrams related by cyclic symmetry, since the contributions in a group can be combined into an integral of the form $\int_0^1 \cdots \prod_{i=1}^{4h} dx_i\delta(\sum_{i=1}^{4h} x_i-1)$,  where the $\cdots$ denotes a product of two complex conjugate factors. One may wonder whether this is also true for the case of three string modes. At genus two, the 21 diagrams are divided into four groups (\ref{biggroups}).  In this case, each group  is closed under the two involution actions $\sigma\rightarrow \sigma^{-1}, \sigma^{\prime}$, so their individual contributions with three string modes are still real and symmetric \footnote{This would be no longer true for higher genus $h\geq 3$. At genus three, there are 1485 diagrams, i.e. permutations of $(1,2,\cdots, 12)$ with 1 fixed at the first position, divided into 131 groups according to cyclic symmetry. Most groups, i.e.  118 of them, have 12 diagrams, while the remaining groups may have 6,4,1 diagrams. There is also a group with a single diagram $(1, 8, 3, 10, 5, 12, 7, 2, 9, 4, 11, 6)$, which is invariant under both $\sigma\rightarrow \sigma^{-1},  \sigma^{\prime}$ actions. Under the action $\sigma\rightarrow \sigma^{-1}$,  27 groups are closed while the others form 52 pairs. Under the action $\sigma\rightarrow \sigma^{\prime}$,  33 groups are closed while the others form 49 pairs. }. We further test the non-negativity for the four groups (\ref{biggroups}) with random mode numbers. In our tests we have not found any negative result for the first three groups. However, for the last group with a single diagram, there are some negative results, though such occurrences are quite rare, e.g. with random mode numbers $m_i, n_i \sim [-30,30] $, about 1/500 of the test results are negative. Of course, the total contributions of the four groups are always non-negative in all our tests.

\section{Conclusion} 	   \label{sectionconclusion}

The main result of the paper is the explicit form (\ref{totalresult}) of BMN two-point functions at genus two. We are able to verify numerically and give an analytic argument in the cases of generic mode numbers with large absolute values that they are always non-negative. Now that we are more confident in the validity of the conjectured non-negativity for three string modes, it would be certainly better to search for a universal analytic proof, at genus two and further at any higher genus. Although the integrals are completely elementary, it seems a complete proof may require some advanced mathematical techniques, probably relating to the moduli space of higher genus Riemann surfaces. Such a proof may elucidate the surprising entry (\ref{newentry}) of the pp-wave holographic dictionary. 
	 
It is well known that string theory has a Hagedorn temperature inverse proportional to the string length, which may be interpreted as a maximal temperature or a temperature of phase transition. In our physical setting, the tensionless strings have effectively infinite length, so the Hagedorn temperature is zero. This is not necessarily a problem, since for example an extremal black hole emits no Hawing radiation, has zero temperature, but can still have finite event horizon and entropy. In \cite{Huang:2019uue}, we consider the entropy of these tensionless BMN strings, which can be interpreted as the Von-Neumann entropy of a mixed BMN state due to decoherence. It would be interesting to see whether we can further study some interesting thermodynamics under such extreme conditions, e.g. along the line of the classic paper \cite{Atick:1988si}. 

It would be very interesting to make connections with the influential Swampland Conjectures \cite{Vafa:2005ui}, which has been studied extensively in the literature, e.g. \cite{Joshi:2019nzi, Kim:2019vuc}. In particular, the Swampland Distance Conjecture \cite{Ooguri:2006in} states that at large distance in the moduli space, an infinite tower of exponentially light states will emerge, invalidate the effective action description.  In our setting, the tensionless strings are an infinite tower of degenerate states with quantum transitions between them without energy cost. Such systems do not have local effective action description but can be still regarded as consistent quantum theories. It would be interesting to better understand the relative place where our present setting fits in the string landscape and the swampland.

It is widely believed that strings may not be the right fundamental degrees of freedom to formulate the still mysterious M-theory, since there are non-perturbative objects like D-branes and M-branes. On the contrary, if our conjecture (\ref{newentry})  is correct, it seems that in our very special setting with infinite spacetime curvature and Ramond-Ramond flux, the tensionless closed strings do provide the proper complete degrees of freedom for physics at any coupling constant. Although we focus on a highly unrealistic special situation, our studies may provide some insights for the non-perturbative formulation of string/M-theory on general backgrounds.

\vspace{0.2in} {\leftline {\bf Acknowledgments}}
\nopagebreak

We thank Jun-Hao Li, Jian-xin Lu, Gao-fu Ren, Pei-xuan Zeng for helpful discussions. This work was supported in parts by the national Natural Science Foundation of China (Grants  No.11947301 and No.12047502).

\appendix

\section{Some Standard Integrals}  \label{standardint}

We will use some standard integrals as in \cite{Du:2021dml}, which is defined by 
\begin{eqnarray} \label{integral1}
I(u_1,u_2,\cdots,u_r) \equiv \int_0^1  dx_1\cdots dx_r \delta(x_1+\cdots+x_r-1) e^{ 2\pi i(u_1x_1+\cdots u_rx_r) }.
\end{eqnarray} 
It is clear that the integral is unchanged if we add an integer to all the arguments. If some of the $u_i$'s are identical, one uses the following notation 
\begin{eqnarray} \label{combine}
I_{(a_1,\cdots,a_r)} (u_1,u_2,\cdots ,u_r)\equiv I(u_1,\cdots, u_1, u_2,\cdots ,u_2, \cdots ,u_r,\cdots ,u_r), 
\end{eqnarray}
where $a_i$'s are integers representing the numbers of the $u_i$'s in the right hand side, and for $a_i=0$ we can just eliminate the corresponding argument. The integral can be calculated by some recursion relations, and one can obtain the following explicit formulas
\begin{eqnarray} \label{integral3}
I(u_1,u_2,\cdots u_r) &=& \sum_{i=1}^r e^{2\pi iu_i} \prod_{j\neq i} \frac{1}{2\pi i(u_i-u_j)} , \\
 \label{integral4}
I_{(a_1+1,\cdots,a_r+1)}(u_1,\cdots,u_r) &=& \prod_{i=1}^r \frac{(\partial / \partial u_i)^{a_i}}{(2\pi i)^{a_i} a_i!} I(u_1,\cdots,u_r),
\end{eqnarray}
where the $u_i$'s are different. We note we have used the $i$ symbol for both the pure imaginary number and the product index, which are easy to distinguish and should not cause confusion. In our calculations, the arguments $u_i$'s will be linear combinations of the integral string mode numbers, so the end results are always rational functions of string mode numbers without the exponential functions. 

A useful special case is when all arguments are degenerate at an integer. Applying the formulas we have 
\be  \label{special}
I_{n+1}(0)  = \int_0^1  dx_1\cdots dx_{n+1} \delta(x_1+\cdots+x_{n+1}-1)  = \frac{1}{n!} ,
\ee
which is simply the volume of the standard $n$-dimensional simplex 
\be 
\Delta_n=\{ (x_1, x_2, \cdots, x_n) \in \mathbb{R}^n | x_i\geq 0, \sum_{i=1}^n x_i\leq 1\}.
\ee 

\addcontentsline{toc}{section}{References}


\providecommand{\href}[2]{#2}\begingroup\raggedright\endgroup

\end{document}